\documentstyle[12pt]{article}
\normalbaselines
\begin{document}
\bibliographystyle{unsrt}

\vbox{\vspace{38mm}}
\begin{center}
{\LARGE \bf Algebraic Realization of
Quark-Diquark Supersymmetry}\\[10mm]

Sultan Catto\\{\it Baruch College and The Graduate School and University Center
\\17 Lexington Avenue, New York, NY 10010}\\
and\\{\it Physics Department, The Rockefeller University\\
1230 York Avenue, New York, NY 10021-6399\\ (e-mail:
cattos@rockvax.rockefeller.edu)}\\[5mm]
 \end{center}

\begin{abstract} Algebraic realizations of supersymmetry through $SU(m/n)$ type 
superalgebras are developed. We show their applications to a bilocal quark-
antiquark or a quark-diquark systems. A new scheme based on $SU(6/1)$ is fully 
exploited and the bilocal approximation is shown to get carried unchanged into 
it. Color algebra based on octonions allows the introduction of a new larger 
algebra that puts quarks, diquarks and exotics in the same supermultiplet as 
hadrons and naturally suppresses quark configurations that are symmetrical in 
color space and antisymmetrical in remaining flavor, spin and position 
variables.  A preliminary work on the first order relativistic formulation 
through the spin realization of Wess-Zumino super-Poincar\'{e} algebra is 
presented.\end{abstract}

\section{Introduction}
To kinds of dynamical supersymmetries are
observed in nature. The first observation came in nuclear physics 
developed by Iachello \cite{ia}, and also by Balantekin, Bars
and Iachello \cite{ibb}. Second observation came in hadronic physics through
an extension of Miyazawa's supersymmetric generalization 
of the approximate $SU(6)$ symmetry of hadrons, namely $SU(6/21)$ that 
was shown to arise within the frame of the standard theory of quarks 
interacting through gluons, built by G\"{u}rsey and Catto \cite{catg},\cite{GS}. Here, 
the 
supersymmetry was found to hold in the same approximation for which $SU(6)$ is 
a good symmetry, provided a diquark structure well separated from the 
remaining quark emerges inside baryons through an effective string 
interaction. At this separation the scalar, spin independent confining part of 
the effective QCD potential is dominant. Since QCD forces are also flavor 
independent, the force between the quark ($q$) and the diquark ($D=qq$) inside 
an excited baryon is essentially the same as the one between $q$ and the 
antiquark $\bar{q}$ inside an excited meson. Thus, the approximate spin-flavor 
independence of hadronic physics gets extended to $SU(6/21)$ supersymmetry 
between the $\bar{q}$ and $D$, resulting into the parallelism of mesonic and 
baryonic Regge trajectories. The $SU(6/21)$ superalgebra for hadrons shown 
below is then applied to a bilocal quark-antiquark $(q-\bar{q})$ or quark-
diquark $(q-D)$ system resulting in multiplets that include the $(\ell = 1)$ 
mesons in the $(35+1)$ representation of $SU(6)$ and the $(\ell = 1)$ baryons 
$(70^{-})$, in addition to the $\ell = 0$ algebra that brings together the 
$(35^{-} + 1^{-})$ mesons and $(56^{+})$ baryons. Color is incorporated through 
the algebra of split octonions.

A relativistic effective Hamiltonian of this quark model was built earlier 
\cite{catg2}, and new mass formulas were derived \cite{catg} \cite{scpu}. This 
Hamiltonian exhibited an approximate $SU(6/21)$ symmetry which was broken by 
the mass differences of the constituents and the spin dependent forces caused 
by gluon exchange. Linear and quadratic mass relationships were derived 
relating the baryon and meson mass differences true to within one percent of 
the experiment.

In this paper we show the algebraic realization of supersymmetry through 
$SU(m/n)$ type superalgebras. A new kind of supersymmetry scheme based on 
octonions is described. These algebras are shown to be embedded in a larger 
octonionic algebra which puts mesons and baryons, exotics, quarks and diquarks 
in the same multiplet. Existence of minimal schemes and properties of the 
algebraic treatment of the symmetries of three quark system is being published 
elsewhere.   

The last section will deal with some preliminary work on the possible 
first order relativistic formulation through the spin realization of the 
Wess-Zumino algebra.

\section{Algebraic Realization of Supersymmetry}                       
                       
   The Miyazawa supersymmetry \cite{mi} based on the supergroup                       
$U(6/21)$ acts on a                        
quark and antidiquark situated at the same point $x_{1}$.  At the                       
point $x_{2}$ we can consider                        
the action of a Miyazawa supergroup with the same parameters, or one with                       
different parameters.  In the first case we have a global symmetry.  In the                        
second case, if we only deal with bilocal fields the symmetry will be                        
represented by $U(6/21)\times U(6/21)$, doubling the Miyazawa supergroup.                        
On the other hand, if any number of points are considered, with                       
different parameters                        
attached to each point, we are led to introduce a local supersymmetry                       
$U(6/21)$ to which we should add the local color group $SU(3)^{c}$.                        
Since it is not a                        
fundamental symmetry, we shall not deal with the local Miyazawa group here.                         
However, the double Miyazawa supergroup is useful for bilocal fields since the                        
decomposition of the adjoint representation of the 728-dimensional Miyazawa                        
group with respect to $SU(6)\times SU(21)$ gives                       
\begin{equation}                      
728= (35,1)+(1,440)+(6,21)+ (\bar{6}, \bar{21}) +(1,1)                      
\end{equation}                       
                       
A further decomposition of the double Miyazawa supergroup into its field with                        
respect to its c.m. coordinates, as will be seen below                       
(see equation (\ref{eq:fat})), leads                        
to the decomposition of the 126-dimensional cosets $(6,21)$ and                       
$(21,6)$ into $56^{+} + 70^{-}$  of the diagonal $SU(6)$.                      
                       
   We would have a much tighter and more elegant scheme if we could perform                        
such a decomposition from the start and be able to identify $(1,21)$                       
part of the fundametal representation of $U(6/21)$ with the $21$-dimensional                       
representation of the $SU(6)$ subgroup, which means going beyond the                       
Miyazawa supersymmetry to a                        
smaller supergroup having $SU(6)$ as a subgroup.  Following two sections will                      
be devoted to the development of a bilocal treatment and a tighter new scheme
based on $SU(3)^{c} \times SU(6/1)$                       
where the bilocal treatment gets carried over unchanged into it.

\section{Bilocal approximation to hadronic structure and inclusion of color.}                       
                       
   Low-lying baryons occur in the symmetric $56$ representation \cite{gr}                       
of $SU(6)$,                        
whereas the Pauli principle would have led to the antisymmetrical $20$                        
representation.  This was a crucial fact for the introduction of color degree                        
of freedom based on $SU(3)^{c}$ \cite{gre}. Since the quark field                       
transforms like a color triplet and the diquark like a color                       
antitriplet under $SU(3)^{c}$, the color                        
degrees of freedom of the constituents must be included correctly in order                        
to obtain a correct representation of the q-D system.  Hadronic states must                        
be color singlets.  These are represented by bilocal operators                       
$O({\bf r}_{1},{\bf r}_{2})$ in the bilocal approximation \cite{tak}                       
that gives $\bar{q}(1)q(2)$ for mesons and $D(1) q(2)$ for                        
baryons.  Here $q(1)$ represents the antiquark situated at ${\bf r}_{1}$,                       
$q(2)$ the quark situated at ${\bf r}_{2}$, and $D(1) = q(1)q(1)$ the                       
diquark situated at ${\bf r}_{1}$.  If we denote                        
the c.m. and the relative coordinates of the consituents by ${\bf R}$ and                       
${\bf r}$, where ${\bf r} ={\bf r}_{2}-{\bf r}_{1}$ and                       
\begin{equation}                      
{\bf R} = \frac{(m_{1} {\bf r}_{1}+m_{2} {\bf r}_{2})}{(m_{1}+m_{2})}                      
\end{equation}                       
with $m_{1}$ and $m_{2}$ being their masses, we can                        
then write $O({\bf R},{\bf r})$ for the operator that creates hadrons                       
out of the vacuum.                         
The matrix element of this operator between the vacuum and the hadronic state                        
$h$ will be of the form                        
\begin{equation}                      
<h| O({\bf R},{\bf r}) |0>= \chi({\bf R}) \psi({\bf r})                      
\end{equation}                       
where $\chi ({\bf R})$ is the free wave function of the hadron as a                       
function of the c.m. coordinate and $\psi ({\bf r})$ is the bound-state                       
solution of the $U(6/21)$ invariant Hamiltonian describing the                       
$q-\bar{q}$ mesons, $q-D$ baryons, $\bar{q}-\bar{D}$ antibaryons and                           
$D-\bar{D}$ exotic mesons, given by                       
\begin{equation}                      
i \partial_{t} \psi_{\alpha \beta} =                       
[\sqrt{(m_{\alpha} + \frac{1}{2} V_{s})^{2} + {\bf p}^{2}} +                      
\sqrt{(m_{\beta} + \frac{1}{2} V_{s})^{2} + {\bf p}^{2}} -\frac{4}{3}                      
\frac{\alpha_{s}}{r} + k \frac{{\bf s}_{\alpha} \cdot{\bf s}_{\beta}}                      
{m_{\alpha} m_{\beta}}] \psi_{\alpha \beta}                      
\end{equation}

Here ${\bf p}= -i \mbox{\boldmath $\nabla$}$ in the c.m. system and $m$ and $s$ denote           
           
the masses and spins of the constituents, $\alpha_{s}$ the                       
strong-coupling constant, $V_{s} = br$ is the scalar potential with $r$                       
being the distance between the constituents in the bilocal object, and                        
$k= |\psi (0)|^{2}$.                      
                       
 The operator product expansion \cite{wil} will give a singular part depending only                       

on ${\bf r}$ and proportional to the propagator of the field binding the two                        
constituents.  There will be a finite number of singular coefficients $c_{n}                      
({\bf r})$ depending on the dimensionality of the constituent fields.                        
For example, for a                        
meson, the singular term is proportional to the progagator of the gluon field                        
binding the two constituents.  Once we subtract the singular part, the                        
remaining part $\tilde{O}({\bf R}, {\bf r})$ is analytic in r and thus                       
we can write                       
\begin{equation}                      
\tilde{O}({\bf R}, {\bf r})= O_{0}({\bf R}) + {\bf r} \cdot{\bf O}_{1}({\bf R})                      
+o(r^{2}).                      
\end{equation}                       
                       
   Now $O_{0}({\bf R})$ creates a hadron at its c.m. point ${\bf R}$                       
equivalent to a $\ell=0$, s-state of                        
the two constituents.  For a baryon this is a state associated with $q$ and                       
$D \sim qq$ at the same point ${\bf R}$, hence it is essentially a 3-quark                       
state when the three quarks are at a common location. The                        
${\bf O}_{1}({\bf R})$ can create three $\ell=1$ states with                        
opposite parity to the state created by $O_{0}({\bf R})$.  Hence, if                       
we denote the nonsingular parts of $\bar{q}(1)q(2)$ and $D(1)q(2)$ by                       
$[\bar{q}(1)q(2)]$ and $[D(1)q(2)]$,respectively, we have                       
\begin{equation}                      
[\bar{q}(1)q(2)]|0> = |M({\bf R})> + {\bf r} \cdot |{\bf M}^{'}({\bf R})> +                      
o(r^{2}) ,                      
\end{equation}                       
\begin{equation}                      
[D(1)q(2)]|0> = |B({\bf R})> + {\bf r} \cdot |{\bf B}^{'}({\bf R})> +                      
o(r^{2}) ,                      
\end{equation}                       
and similarly for the exotic meson states $D(1) \bar{D}(2)$.                      
                       
   Here M belongs to the $(35+1)$-dimensional representation of $SU(6)$                        
corresponding to $\ell=0$ bound state of the quark and the antiquark.                        
The $M^{'}({\bf R})$ is an orbital excitation $(\ell=1)$ of opposite                       
parity, which are in the $(35+1,3)$ representation of the group                       
$SU(6) \times O(3)$, $O(3)$ being associated with the relative                        
angular momentum of the constituents.  The $M^{'}$ states contain                       
mesons like $B$, ${\bf A}_{1}$, ${\bf A}_{2}$ and scalar particles.                        
On the whole, the $\ell=0$ and $\ell=1$ part $\bar{q}(1)q(2)$ contain                       
$4 \times(35+1) = 144$ meson states.                      
                       
   Switching to the baryon states, the requirement of antisymmetry in color,                       
and symmetry in spin-flavor indices gives the $(56)^{+}$                       
representation for $B({\bf R})$. The $\ell=1$ multiplets have negative                       
parity and have mixed spin-flavor symmetry.                       
They belong to the representation $(70^{-},3)$ of $SU(6) \times O(3)$ and                       
are represented by the states $|{\bf B}^{'}({\bf R})>$ which are $210$ in                       
number.  On the whole, these $266$ states                       
account for all the observed low-lying baryon states obtained form                        
$56 + 3 \times 70 = 266$.  A similar analysis can be carried out for the                       
exotic meson states $D(1) \bar{D}(2)$, where the diquark and the                       
antidiquark can be bound in a $\ell=0$ or $\ell=1$ state with                       
opposite parities.

\section{Color algebra and Octonions.}                       
                       
   The behavior of various states under the color group are best                       
seen if we use split octonion units defined by \cite{gug}                       
\begin{equation}                      
u_{0} = \frac{1}{2} (1 +i e_{7}) ,                      
~~~~~~~u_{0}^{*} = \frac{1}{2} (1 -i e_{7}) ,                      
\end{equation}                        
\begin{equation}                      
u_{j} = \frac{1}{2} (e_{j} +i e_{j+3}) ,                      
~~~~~u_{j}^{*} = \frac{1}{2} (e_{j} -i e_{j+3}) , ~~~j=1,2,3.                      
\end{equation}                       
                      
The automorphism group of the octonion algebra is the 14-parameter                       
exceptional group $G_{2}$.  The imaginary octonion units                       
$e_{\alpha} (\alpha  =1,...,7)$                       
fall into its 7-dimensional representation.                      
                       
   Under the $SU(3)^{c}$ subgroup of $G_{2}$ that leaves $e_{7}$                       
invariant, $u_{0}$ and $u_{0}^{*}$ are singlets, while $u_{j}$ and                       
$u_{j}^{*}$ correspond, respectively, to the                       
representations ${\bf 3}$ and $\bar{\bf 3}$.  The multiplication table can now be                       
written in a manifestly $SU(3)^{c}$ invariant manner (together with the                       
complex conjugate equations):                       
\begin{equation}                      
u_{0}^{2} = u_{0},~~~~~u_{0}u_{0}^{*} = 0                      
\end{equation}                       
\begin{equation}                      
u_{0} u_{j} = u_{j} u_{0}^{*} = u_{j},~~~~~                      
u_{0}^{*} u_{j} = u_{j} u_{0} = 0 ,                         
\end{equation}                       
\begin{equation}                      
u_{i} u_{j}  = - u_{j} u_{i} = \epsilon_{ijk} u_{k}^{*},      \label{eq:oct}                      
\end{equation}                       
\begin{equation}                      
u_{i} u_{j}^{*} =  - \delta_{ij} u_{0}         \label{eq:octa}                      
\end{equation}                       
where $\epsilon_{ijk}$ is completely antisymmetric with  $\epsilon_{ijk} =1$                      
for  $ijk$  = $123$, $246$, $435$, $651$, $572$, $714$, $367$.                        
Here, one sees the                       
virtue of octonion multiplication.  If we consider the direct products                       
\begin{equation}                      
C:~~~~~{\bf 3} \otimes \bar{\bf 3} = {\bf 1} + {\bf 8} ,                      
\end{equation}                       
\begin{equation}                      
G:~~~~~{\bf 3} \otimes {\bf 3} = \bar{\bf 3} + {\bf 6}                      
\end{equation}                       
for $SU(3)^{c}$, then these equations show that octonion multiplication                       
gets rid of ${\bf 8}$ in ${\bf 3} \otimes \bar{\bf 3}$, while it gets rid                       
of ${\bf 6}$ in ${\bf 3} \otimes {\bf 3}$.  Combining  (\ref{eq:oct}) and                      
(\ref{eq:octa}) we find                       
\begin{equation}                      
(u_{i} u_{j}) u_{k} = - \epsilon_{ijk} u_{0}^{*} .                      
\end{equation}                       
                       
   Thus the octonion product leaves only the color part in                       
${\bf 3} \otimes \bar{\bf 3}$ and ${\bf 3} \otimes {\bf 3} \otimes {\bf 3}$,                      
so that it is a natural algebra for colored quarks.                       
                      
   The quarks, being in the triplet representation of the color                       
group $SU(3)^{c}$, they are represented by the local fields                      
$q_{\alpha}^{i}(x)$, where $i = 1,2,3$ is the color index and $\alpha$                        
the combined spin-flavor index. Antiquarks at point $y$ are color                       
antitriplets $q_{\beta}^{i}(y)$.  Condider the two-body systems                       
\begin{equation}                      
C_{\alpha j}^{\beta i} = q_{\alpha}^{i} (x_{1}) \bar{q}_{\beta}^{j} (x_{2}) ,                      
\label{eq:c}                      
\end{equation}                       
\begin{equation}                      
G_{\alpha \beta}^{i j} = q_{\alpha}^{i} (x_{1}) q_{\beta}^{j} (x_{2}) ,                      
\label{eq:g}                      
\end{equation}                       
so that $C$ is either a color siglet or color octet, while $G$ is a                       
color antitriplet or a color sextet.  Now $C$ contains meson states                       
that are color singlets and hence observable.  The octet $q-\bar{q}$ state                       
is confined and not observed as a scattering state.  In the case of                       
two-body $G$ states, the antitriplets are diquarks which, inside a                       
hadron can be combined with another triplet quark to give                       
observable, color singlet, three-quark baryon states.  The color                       
sextet part of $G$ can only combine with a third quark to give                       
unobservable color octet and color decuplet three-quark states.                        
Hence the hadron dynamics is such that the ${\bf 8}$ part of $C$                       
and the ${\bf 6}$                       
part of $G$ are suppressed.  This can best be achieved by the use of                       
above octonion algebra \cite{dom}.  The dynamical suppression of the                       
octet and sextet states in (\ref{eq:c}) and (\ref{eq:g}) is , therefore,                       
automatically achieved.  The split octonion units can be contracted                       
with color indices of triplet or antitriplet fields.  For quarks                       
and antiquarks we can define the "transverse" octonions (calling $u_{0}$                       
and $u_{0}^{*}$ longitidunal units)                       
\begin{equation}                      
q_{\alpha} = u_{i} q_{\alpha}^{i} = {\bf u} \cdot{\bf q}_{\alpha} ,~~~~~                      
\bar{q}_{\beta} = u_{i}^{\dagger} \bar{q}_{\beta}^{j} = -{\bf u}^{*}                       
\cdot \bar{\bf q}_{\beta} .       \label{eq:qal}                      
\end{equation}                       
                       
We find                       
\begin{equation}                      
q_{\alpha}(1) \bar{q}_{\beta}(2) = u_{0} {\bf q}_{\alpha}(1)                       
\cdot{\bf q}_{\beta}(2),                      
\end{equation}                       
\begin{equation}                      
\bar{q}_{\alpha}(1) q_{\beta}(2) = u_{0}^{*} \bar{\bf q}_{\alpha}(1)                       
\cdot{\bf q}_{\beta}(2),                      
\end{equation}                       
\begin{equation}                      
G_{\alpha \beta}(12) = q_{\alpha}(1) q_{\beta}(2) = {\bf u}^{*}                       
\cdot{\bf q}_{\alpha}(1) \times {\bf q}_{\beta}(2),                      
\end{equation}                       
\begin{equation}                      
G_{\beta \alpha}(21) = q_{\beta}(2) q_{\alpha}(1) = {\bf u}^{*}                       
\cdot{\bf q}_{\beta}(2) \times {\bf q}_{\alpha}(1).                      
\end{equation}                       
                       
Because of the anticomutativity of the quark fields, we have                       
\begin{equation}                      
G_{\alpha \beta}(12) = G_{\beta \alpha}(21) =                       
\frac{1}{2}  \{q_{\alpha}(1), q_{\beta}(2)\}.                      
\end{equation}                       
                       
If the diquark forms a bound state represented by a field $D_{\alpha \beta}(x)$                      
at the center-of-mass location $x$                       
\begin{equation}                      
x = \frac{1}{2} (x_{1} +x_{2}) ,                      
\end{equation}                       
when $x_{2}$ tends to $x_{1}$ we can replace the argument by $x$, and we obtain                      

\begin{equation}                      
D_{\alpha \beta}(x) = D_{\beta \alpha}(x) ,                      
\end{equation}                       
so that the local diquark field must be in a symmetric                       
representation of the spin-flavor group.  If the latter is taken to                       
be $SU(6)$, then $D_{\alpha \beta}(x)$ is in the 21-dimensional symmetric                       
representation, given by                        
\begin{equation}                      
({\bf 6} \otimes {\bf 6})_{s} = {\bf 21} .                      
\end{equation}                       
                       
If we denote the antisymmetric $15$ representation by $\Delta_{\alpha \beta}$,                      
we see that the octonionic fields single out the $21$ diquark representation                       
at the expense of $\Delta_{\alpha \beta}$.  We note that without this                       
color algebra supersymmetry would give antisymmetric configurations as noted by                       
Salam and Strathdee \cite{sal} in their possible supersymmetric                       
generalization of hadronic supersymmetry.  Using the nonsingular                       
part of the operator product expansion we can write                       
\begin{equation}                      
\tilde{G}_{\alpha \beta}({\bf x}_{1}, {\bf x}_{2}) =                      
D_{\alpha \beta}(x) + {\bf r} \cdot {\bf \Delta}_{\alpha \beta}(x).                      
\label{eq:qam}                      
\end{equation}                       
The fields $\Delta_{\alpha \beta}$ have opposite parity to $D_{\alpha \beta}$;                      
${\bf r}$ is the relative                       
coordinate at time $t$ if we take $t$ = $t_{1}$ = $t_{2}$.  They play no role in                       
the excited baryon which becomes a bilocal system with the 21-                       
dimensional diquark as one of its constituents.                       
                      
   Now consider a three-quark system at time $t$.  The c.m. and                       
relative coordinates are                       
\begin{equation}                      
{\bf R} = \frac{1}{\sqrt{3}}({\bf r}_{1} + {\bf r}_{2} + {\bf r}_{3}),                      
\end{equation}                       
\begin{equation}                      
\mbox{\boldmath $\rho$} = \frac{1}{\sqrt{6}}(2 {\bf r}_{3} - {\bf r}_{1} - {\bf r}_{2}),            
\end{equation}                       
\begin{equation}                      
{\bf r} = \frac{1}{\sqrt{2}}({\bf r}_{1} - {\bf r}_{2}),                      
\end{equation}                        
giving                       
\begin{equation}                      
{\bf r}_{1} = \frac{1}{\sqrt{3}} {\bf R} - \frac{1}{\sqrt{6}} 
\mbox{\boldmath $\rho$}                      
+ \frac{1}{\sqrt{2}} {\bf r}                      
\end{equation}                       
\begin{equation}                      
{\bf r}_{2} = \frac{1}{\sqrt{3}} {\bf R} - \frac{1}{\sqrt{6}} 
\mbox{\boldmath $\rho$}                      
- \frac{1}{\sqrt{2}} {\bf r}                      
\end{equation}                       
\begin{equation}                      
{\bf r}_{3} = \frac{1}{\sqrt{3}} {\bf R} + \frac{2}{\sqrt{6}} 
\mbox{\boldmath $\rho$}                      
\end{equation}                       
                       
The baryon state must be a color singlet, symmetric in the three                       
pairs ($\alpha$, $x_{1}$), ($\beta$, $x_{2}$), ($\gamma$, $x_{3}$).  We find                       
\begin{equation}                      
(q_{\alpha}(1) q_{\beta}(2)) q_{\gamma}(3) = -u_{0}^{*}                       
F_{\alpha \beta \gamma}(123) ,                      
\end{equation}                        
\begin{equation}                      
q_{\gamma}(3) (q_{\alpha}(1) q_{\beta}(2))  = -u_{0}                       
F_{\alpha \beta \gamma}(123) ,                      
\end{equation}                        
so that                       
\begin{equation}                      
- \frac{1}{2} \{\{q_{\alpha}(1), q_{\beta}(2)\}, q_{\gamma}(3)\} =                        
F_{\alpha \beta \gamma}(123) .                      
\end{equation}                       
                       
The operator $F_{\alpha \beta \gamma}(123)$ is a color singlet and is                       
symmetrical in the three pairs of coordinates.  We have                       
\begin{equation}                      
F_{\alpha \beta \gamma}(123) = B_{\alpha \beta \gamma} ({\bf R}) +                      
\mbox{\boldmath $\rho$} \cdot {\bf B}'({\bf R}) + {\bf r} \cdot 
{\bf B}"({\bf R}) + C                      
\label{eq:fat}                      
\end{equation}                      
where $C$ is of order two and higher in $\mbox{\boldmath $\rho$}$ and 
${\bf r}$.  Because                       
${\bf R}$ is symmetric in ${\bf r}_{1}$, ${\bf r}_{2}$ and ${\bf r}_{3}$,                       
the operator $B_{\alpha \beta \gamma}$    that creates a baryon                       
at ${\bf R}$ is totally symmetrical in its flavor-spin indices.  In the                       
$SU(6)$ scheme it belongs to the ($56$) representation.  In the bilocal                       
$q-D$ approximation we have ${\bf r}=0$ so that $F_{\alpha \beta \gamma}$ is                       
a function only of ${\bf R}$ and $\mbox{\boldmath $\rho$}$ which are both 
symmetrical in                       
${\bf r}_{1}$ and ${\bf r}_{2}$.  As before, ${\bf B}'$                        
belongs to the orbitally excited $70^{-}$ represenation of $SU(6)$.  The                       
totally antisymmetrical ($20$) is absent in the bilocal                       
approximation.  It would only appear in the trilocal treatment that                       
would involve the 15-dimensional diquarks.  Hence, if we use local                       
fields, any product of two octonionic quark fields gives a ($21$)                       
diquark                       
\begin{equation}                      
q_{\alpha}({\bf R}) q_{\beta}({\bf R}) = D_{\alpha \beta}({\bf R}),                      
\end{equation}                       
and any nonassociative combination of three quarks, or a diquark                       
and a quark at the same point give a baryon in the $56^{+}$ representation:                       
\begin{equation}                      
(q_{\alpha}({\bf R}) q_{\beta}({\bf R})) q_{\gamma}({\bf R}) = - u_{0}^{*}                       
B_{\alpha \beta \gamma}({\bf R}),                      
\end{equation}                       
\begin{equation}                      
q_{\alpha}({\bf R}) (q_{\beta}({\bf R}) q_{\gamma}({\bf R})) = - u_{0}                       
B_{\alpha \beta \gamma}({\bf R}),                      
\end{equation}                       
\begin{equation}                      
q_{\gamma}({\bf R}) (q_{\alpha}({\bf R}) q_{\beta}({\bf R})) = - u_{0}                       
B_{\alpha \beta \gamma}({\bf R}),                      
\end{equation}                       
\begin{equation}                      
(q_{\gamma}({\bf R}) q_{\alpha}({\bf R})) q_{\beta}({\bf R}) = - u_{0}^{*}                       
B_{\alpha \beta \gamma}({\bf R}).                      
\end{equation}                       
                       
The bilocal approximation gives the ($35+1$) mesons and the $70^{-}$                       
baryons with $\ell=1$ orbital excitation.

\section{A colored supersymmetry scheme based on $SU(3)^{c} \times SU(6/1)$}                    
                    
We could  go to a smaller supergroup having $SU(6)$ as a subgroup. With the addition of color,
such a supergroup is $SU(3) \times SU(6/1)$. The fundamental representation of $SU(6/1)$ is
$7$-dimensional which decomposes into a sextet and singlet under the spin-flavor group. There
is also a $28$-dimensional representation of $SU(7)$. Under the $SU(6)$ subgroup it has the
decomposition                    
\begin{equation}                    
28 = 21 + 6 + 1  .                  
\end{equation}                    
                    
Hence, this supermultiplet can accommodate the bosonic antidiquark and 
fermionic quark in it, provided we are willing to add another scalar. Together 
with the color symmetry, we are led to consider the ($3$, $28$) representation 
of $SU(3) \times SU(6/1)$ which consists of an antidiquark, a quark and a color 
triplet scalar that we shall call a scalar quark. This boson is in some way 
analogous to the $s$ quarks. The whole multiplet can be represented by an 
octonionic $7 \times 7$ matrix $Z$ at point $x$.                    
                    
\begin{equation}                    
Z = {\bf u} \cdot                    
\left(                    
\begin{array}{cc}                    
\overline{\bf D}^{*} & {\bf q} \\                    
i {\bf q}^{T}  \sigma_{2} & {\bf S}                    
\end{array} \right)                    
\end{equation}                    
Here ${\bf D}^{*}$ is a $6 \times 6$ symmetric matrix representing the 
antidiquark, 
${\bf q}$ is a $6 \times 1$ column matrix,             
${\bf q}^{T}$ is its transpose and $\sigma_{2}$  is the Pauli matrix that 
acts on the spin indices of the quark so that, if $q$ transforms with the 
$2 \times 2$ Lorentz matrix $L$,                     
${\bf q}^{T} i \sigma_{2}$ transforms with $L^{-1}$ acting from the right.                    
                    
Similarly we have                      
                    
\begin{equation}                    
Z^{c} = {\bf u}^{*} \cdot                     
\left(                    
\begin{array}{cc}                    
{\bf D} & i \sigma_{2} {\bf q}^{*} \\                    
{\bf q}^{\dag} & {\bf S}^{*}                    
\end{array} \right)                    
\end{equation}                    
to represent the supermultiplet with a diquark and antidiquark. The mesons, exotic mesons and
baryons are all in the bilocal field $Z(1) \otimes Z^{c}(2)$ which we expend with respect to the
center of mass coordinates in order to represent color singlet hadrons by local fields. The color
singlets $56^{+}$ and $70^{-}$ will then arise as in the earlier section.                    
                    
Now the ($\bar{D} q$) system belonged to the fundamental representation of the $SU(6/21)$
supergroup. But $Z$ belongs to the ($28$) representation of $SU(6/1)$ which is not its
fundamental representation. Are there any fields that belong to the $7$-dimensional
representation of $SU(6/1)$? It is possible to introduce such fictitious fields a $6$-dimensional
spinor $\xi$ and a scalar $a$ without necessarily assuming their existence as particles. We put

\begin{equation}                    
\xi = {\bf u}^{*} \cdot \mbox{\boldmath $\xi$} , ~~~~~ a = {\bf u}^{*} \cdot 
{\bf a} ,                    
\end{equation}                    
so that both $\mbox{\boldmath $\xi$}$ and ${\bf a}$ are color antitriplets. Let                    
\begin{equation}                    
\lambda = \left(             
\begin{array}{c}             
\xi \\ a             
\end{array}                    
\right) ,~~~~~~                    
\lambda^{c} = \left(             
\begin{array}{c}             
\hat{\xi} \\ a^{*}             
\end{array}                    
\right)                    
\end{equation}                    
where                    
\begin{equation}                 
\hat{\xi} = {\bf u} \cdot (i \sigma_{2} \mbox{\boldmath $\xi$}^{*}) ,~~~~~                 
a^{*} = {\bf u} \cdot {\bf a}^{*} .                 
\end{equation}                 
                 
Concider the $7 \times 7$ matrix                 
\begin{equation}                 
W = \lambda \lambda^{c \dag} = \left(                  
\begin{array}{cc}                 
\xi \hat{\xi}^{\dag} & \xi a \\                 
a   \hat{\xi}^{\dag} & 0                 
\end{array}  \right) .                 
\end{equation}                 
                 
$W$ belongs to the $28$-dimensional representation of $SU(6/1)$ and transforms like $Z$,
provided the components of 
$\mbox{\boldmath $\xi$}$ are Grassmann numbers and ${\bf a}$ are even (bosonic)
coordinates. The identification of $Z$ and $W$ would give                 
\begin{equation}                 
{\bf s} = {\bf a} \times {\bf a} = 0,~~{\bf q}_{\alpha} = \mbox{\boldmath 
$\xi$}_{\alpha} \times {\bf a},~~                 
D_{\alpha \beta}^{*} = \mbox{\boldmath $\xi$}_{\alpha} \times 
\mbox{\boldmath $\xi$}_{\beta}.  \end{equation}               
                 
 A scalar part in $W$ can be generated by multiplying two different $(7)$ representations. The
$56^{+}$ baryons form the color singlet part of the $84$-dimensional representation of
$SU(6/1)$ while its colored part consists of quarks and diquarks.                 
                             
Now consider the octonionic valued quark field $q_{A}^{i}$, where    
$i = 1, 2, 3$ is the color index and $A$ stands for the pair    
$(\alpha, \mu)$ with $\alpha = 1, 2$ being the spin index and $\mu =    
1, 2, 3$ the flavor index. (If we have $N$ flavors, $A = 1, \ldots , N$).    
As before     
    
\begin{equation}    
q_{A} = u_{i} q_{A}^{i} = {\bf u} \cdot{\bf q}_{A} .    
\end{equation}    
Similarly the diquark $D_{AB}$ which transforms like a color antitriplet    
is    
    
\begin{equation}    
D_{AB} = q_{A} q_{B} = q_{B} q_{A} = \epsilon_{ijk} u_{k}^{*}     
q_{A}^{i} q_{B}^{j} = {\bf u}^{*} \cdot{\bf D}_{AB} .    
\end{equation}    
    
We note, once again, that because $q_{A}^{i}$ are anticommuting fermion     
operators, $D_{AB}$ is symmetric in its two indices. The antiquark and     
antidiquark are represented by    
\begin{equation}    
\bar{q}_{A} = {\bf u}^{*} \cdot \bar{{\bf q}}_{A},    
\end{equation}   
and   
\begin{equation}    
\bar{D}_{AB} = {\bf u} \cdot \bar{{\bf D}}_{AB},    
\end{equation}    
respectively. If we have $3$ flavors $q_{A}$ has $6$ components for each    
color while $D_{AB}$ has $21$ components.    
    
At this point let us study the system ($q_{A}$, $\bar{{\bf D}}_{BC}$)    
consisting of a quark and an antidiquark, both color triplets. There are     
two possibilities: we can regard the system as a multiplet belonging to     
the fundamental representation of a supergroup $U(6/21)$ for each color,    
or as a higher representation of a smaller supergroup. The latter possibility     
is more economical. To see what kind of supergroup we can have, we imagine    
that both quarks and diquarks are components of more elementary quantities:     
a triplet fermion $f_{A}^{i}$ and a boson $C^{i}$ which is a triplet    
with respect to the color group and a singlet with respect to $SU(2N)$    
($SU(6)$) for three flavors). The $f_{A}^{i}$ is taken to have baryon     
number $1/3$ while $C$ has baryon number $-2/3$. The system     
\begin{equation}    
q_{A}^{i} = \epsilon_{ijk} \bar{f}_{A}^{j} \bar{C}^{k}    
\end{equation}    
will be a color triplet with baryon number $1/3$. It can therefore     
represent a quark. We can write    
\begin{equation}   
q_{A} = {\bf u} \cdot{\bf q}_{A} =   
( {\bf u}^{*} \cdot \bar{{\bf f}}_{A}) ( {\bf u}^{*} \cdot \bar{{\bf C}})   
\end{equation}   
With two anti-f fields we can form bosons that have same quantum numbers as   
antidiquarks:   
\begin{equation}   
\bar{D}_{AB} = {\bf u} \cdot \bar{{\bf D}}_{AB} =   
( {\bf u}^{*} \cdot \bar{{\bf f}}_{A}) ( {\bf u}^{*} \cdot \bar{{\bf f}}_{B}) .   
\end{equation}   
   
In this case the basic multiplet is $({\bf f}_{A}, {\bf C})$ which belongs    
to the fundamental representation of $SU(6/1)$ for each color component. The   
complete algebra to consider is $SU(3) \times SU(6/1)$ and the basic multiplet    
corresponds to the representation $(3, 7)$ of this algebra. Let   
\begin{equation}   
F =    
\left(   
\begin{array}{c}   
f_{1} \\ f_{2} \\ \vdots \\ f_{6} \\ C   
\end{array} \right),   
~~~~~f_{A} = {\bf u} \cdot {\bf f}_{A}, ~~~~~ C = {\bf u} \cdot {\bf C} .   
\end{equation}   
   
Also let   
\begin{equation}   
f^{1} =    
\left(   
\begin{array}{c}   
f_{1}^{1} \\ f_{2}^{1} \\ \vdots \\ f_{6}^{1}   
\end{array} \right),~~~~~   
f^{2} =    
\left(   
\begin{array}{c}   
f_{1}^{2} \\ f_{2}^{2} \\ \vdots \\ f_{6}^{2}   
\end{array} \right).   
\end{equation}   
   
Combining two such representations and writing $\bar{{\bf X}} = {\bf F} \times   
{\bf F}^{T}$ we have   
\begin{equation}   
\bar{X} = {\bf u}^{*} \cdot \bar{{\bf X}} =   
\left(   
\begin{array}{c}   
f^{1} \\ C^{1}   
\end{array} \right)   
\left(   
\begin{array}{cc}   
f^{2T} &   C^{2}   
\end{array} \right) -   
\left(   
\begin{array}{c}   
f^{2} \\ C^{2}   
\end{array} \right)   
\left(   
\begin{array}{cc}   
f^{1T} &   C^{1}   
\end{array} \right).   
\end{equation}   
   
Further identifying    
\begin{equation}   
{\bf u}^{*} \cdot {\bf D}_{11} = 2 f_{1}^{1} f_{1}^{2}, ~~~~~   
{\bf u}^{*} \cdot {\bf D}_{12} =  f_{1}^{1} f_{2}^{2} -   
f_{1}^{2} f_{2}^{1},~~  etc.,   
\end{equation}   
and   
\begin{equation}   
{\bf u}^{*} \cdot \bar{{\bf q}}_{1} = f_{1}^{1} C^{2} - f_{1}^{2} C^{1},~~~~   
{\bf u}^{*} \cdot \bar{{\bf q}}_{2} = f_{2}^{1} C^{2} - f_{2}^{2} C^{1},~~   
etc.,   
\end{equation}   
we see that $\bar{X}$ has the structure   
\begin{equation}   
\bar{X} = {\bf u}^{*} \cdot \bar{{\bf X}}    
 = {\bf u}^{*} \cdot   
\left(   
\begin{array}{cccc}   
{\bf D}_{11} & \ldots & {\bf D}_{16} & \bar{{\bf q}}_{1} \\   
{\bf D}_{12} & \ldots & {\bf D}_{26} & \bar{{\bf q}}_{2} \\   
{\bf D}_{13} & \ldots & {\bf D}_{36} & \bar{{\bf q}}_{3} \\   
{\bf D}_{14} & \ldots & {\bf D}_{46}  & \bar{{\bf q}}_{4} \\   
{\bf D}_{15} & \ldots & {\bf D}_{56}  & \bar{{\bf q}}_{5} \\   
{\bf D}_{16} & \ldots & {\bf D}_{66}  & \bar{{\bf q}}_{6} \\   
-\bar{{\bf q}}_{1} & \ldots & -\bar{{\bf q}}_{6} &  0            
\end{array} \right)   
\end{equation}   
or   
\begin{equation}   
( 3, 7) \times ( 3, 7) = (\bar{3} \times 27 )   
\end{equation}   
   
The $27$ dimensional representation decomposes into $21 + \bar{6}$ with   
respect to its $SU(6)$ subgroup.   
   
Consider now an antiquark-diquark system at point $x_{1} = x -   
\frac{1}{2} \xi$ and another quark-antidiquark system at point    
$x_{2} = x + \frac{1}{2} \xi$; Hence we take the direct product of    
$X(x_{1})$ and $X(x_{2})$. In other words   
\begin{equation}   
(D_{AB} (x_{1}), \bar{q}_{D} (x_{1})) \otimes   
(\bar{D}_{EF} (x_{2}), q_{C} (x_{2}))    
\end{equation}   
consisting of the pieces    
\begin{equation}   
H(x_{1}, x_{2}) = \left(   
\begin{array}{cc}   
\bar{q}_{D}(x_{1}) q_{C}(x_{2}) & D_{AB}(x_{1}) q_{C}(x_{2}) \\   
\bar{q}_{D}(x_{1}) \bar{D}_{EF}(x_{2}) & D_{AB}(x_{1}) \bar{D}_{EF}(x_{2})   
\end{array}  \right)   
\end{equation}   
   
The diagonal pieces are bilocal fields representing color singlet   
$1 + 35$ mesons and $1 + 35 + 405$ exotic mesons respectively with respect    
to the subgroup $SU(3)^{c} \times SU(6)$ of the algebra. The off diagonal   
pieces are color singlets that are completely symmetrical with respect to   
the indices $(ABC)$ and $(DEF)$. They correspond to baryons and antibaryons   
in the representations $56$ and $\bar{56}$ respectively of $SU(6)$.   
   
We can write $F_{ABC} = - \frac{1}{2} \{D_{AB}, q_{C}\}$ so that   
\begin{eqnarray}   
D_{AB} q_{C}&= & (u_{1}^{*} (q_{A}^{2} q_{B}^{3} + q_{B}^{2} q_{A}^{3}) +   
u_{2}^{*} (q_{A}^{3} q_{B}^{1} + q_{B}^{3} q_{A}^{1}) +   
u_{3}^{*} (q_{A}^{1} q_{B}^{2} + q_{B}^{1} q_{A}^{2})) \times  \nonumber  \\
            &  & 
(u_{1} q_{C}^{1} + u_{2} q_{C}^{2} + u_{3} q_{C}^{3}) ,   
\end{eqnarray}   
becomes   
\begin{equation}   
D_{AB} q_{C}=    
- u_{0}^{*} ( (q_{A}^{2} q_{B}^{3} + q_{B}^{2} q_{A}^{3}) q_{C}^{1} +   
(q_{A}^{3} q_{B}^{1} + q_{B}^{3} q_{A}^{1}) q_{C}^{2} +   
(q_{A}^{1} q_{B}^{2} + q_{B}^{1} q_{A}^{2}) q_{C}^{3}) ,   
\end{equation}   
and similarly   
\begin{equation}   
q_{C} D_{AB} =    
- u_{0}  ((q_{A}^{2} q_{B}^{3} + q_{B}^{2} q_{A}^{3}) q_{C}^{1} +   
(q_{A}^{3} q_{B}^{1} + q_{B}^{3} q_{A}^{1}) q_{C}^{2} +   
(q_{A}^{1} q_{B}^{2} + q_{B}^{1} q_{A}^{2}) q_{C}^{3}).   
\end{equation}   
Since $u_{0} + u_{0}^{*} = 1$, we have   
\begin{eqnarray}   
F_{ABC} = - \frac{1}{2} \{D_{AB}, q_{C}\}& = &  
(q_{A}^{1} q_{B}^{2} + q_{B}^{1} q_{A}^{2}) q_{C}^{3} +   
(q_{A}^{2} q_{B}^{3} + q_{B}^{2} q_{A}^{3}) q_{C}^{1} +  \nonumber   \\
                                         &   &
(q_{A}^{3} q_{B}^{1} + q_{B}^{3} q_{A}^{1}) q_{C}^{2}    
\end{eqnarray}   
which is completely symmetric with respect to indices $(ABC)$, corresponding    
to baryons.    
   
In the limit $x_{2} - x_{1} = \xi \longrightarrow 0$, $H$ can be represented   
by a local supermultiplet with dimension $ 2 \times 56 + 2(1 + 35) + 405 =   
589$ of the original algebra. This representation includes $56$ baryons,   
antibaryons, mesons and $q^{2} \bar{q}^{2}$ exotic mesons.   
   
\section{Transformation properties}   
   
Since $F = {\bf u} \cdot {\bf F}$ consists of three $7$-dimensional   
representations of $SU(6/1)$ we have   
\begin{equation}   
\delta F = Z~~F   
\end{equation}   
where $Z \in SU(6/1)$,   
\begin{equation}    
F =    
\left(   
\begin{array}{c}   
f \\ C   
\end{array} \right)   
=   
\left(   
\begin{array}{c}   
f_{1} \\ f_{2} \\ \vdots \\ f_{6} \\ C   
\end{array} \right) .   
\end{equation}   
Hence $Z$ is super antihermitian color singlet:   
\begin{equation}   
Z = \left(   
\begin{array}{cc}   
i H & \eta \\   
i \eta^{\dag} & i \omega   
\end{array}  \right) ,   
\end{equation}   
with $ H = H^{\dag}$, $ \omega = \omega^{*} $, $ Str Z = 0 $    
($Str$ =    
supertrace), $(\omega = tr H)$, and     
\begin{equation}   
\eta =    
\left(   
\begin{array}{c}   
\eta_{1} \\ \eta_{2} \\ \vdots \\ \eta_{6}   
\end{array} \right),   
~~~~~ \{ \eta_{\alpha}, \eta_{\beta} \} = 0 .   
\end{equation}   
$H$ is an antihermitian $6 \times 6$ matrix.    
   
Then by taking supertransposed quantities ($sT$ = supertransposed)   
\begin{equation}   
\delta F^{sT} = \delta F^{T} = F^{T}~Z^{sT}   
\end{equation}   
we have   
\begin{equation}   
\left(   
\begin{array}{c}   
\delta {\bf f} \\  \delta {\bf C}   
\end{array} \right)   
=    
 \left(   
\begin{array}{cc}   
i H & \eta \\   
i \eta^{\dag} & i \omega   
\end{array}  \right)   
\left(   
\begin{array}{c}   
{\bf f} \\  {\bf C}   
\end{array} \right) ,   
\end{equation}   
\begin{equation}   
\left(   
\begin{array}{cc}   
\delta {\bf f}^{T} &  \delta {\bf C}   
\end{array} \right) =   
 \left(   
\begin{array}{cc}   
i H^{*} & -i \eta^{*} \\   
 \eta^{T} & i \omega   
\end{array}  \right) .   
\end{equation}   
We have   
\begin{equation}   
\bar{X} = F~~F^{T} = F~~F^{sT}   
\end{equation}   
so that   
\begin{equation}   
\delta \bar{X} = Z~~\bar{X} + \bar{X}~~Z^{sT}.   
\end{equation}   
   
Writing   
\begin{equation}   
\bar{X} =   
 \left(   
\begin{array}{cc}   
\bar{D} & q \\   
 - q^{T} & 0   
\end{array}  \right) ,    
~~~~~~(D = D^{T}) .   
\end{equation}   
under $U(6/21)$, $\delta X$ gives:   
\begin{equation}   
\delta \bar{D} = i (H \bar{D} + \bar{D} H^{T}) - (\eta q^{T} - q \eta^{T}) ,   
\end{equation}   
\begin{equation}   
\delta q = i ( H + \omega ) q - \bar{D} \eta^{*} .   
\end{equation}   
For supertransformation $SU(6/1)~/~U(6)$, the change in $\bar{D}$ and $q$ are   
   
\begin{equation}   
\delta \bar{D}=   q \eta^{T} - \eta q^{T}  ,~~~~~ \delta q =   
- \bar{D} \eta^{*} .   
\end{equation}   
   
Since   
\begin{equation}   
\delta f = Z f ,   
\end{equation}   
or in component notation   
\begin{equation}   
\delta f_{A} = Z_{AB} f_{B}   
\end{equation}   
where $(A,B = 0, 1, \ldots, 6)$ with $f_{0} = C$, we have    
\begin{equation}   
\delta f^{sT*} = \delta f^{\dag} = f^{\dag}~Z^{*sT} .   
\end{equation}   
   
If we define $g$ by   
\begin{equation}   
g =    
 \left(   
\begin{array}{cc}   
I  & 0 \\   
 0 &  i   
\end{array}  \right) ,    
\end{equation}   
then   
\begin{equation}   
\delta (f^{\dag}~g~f) =    
f^{\dag} (g Z + Z^{*sT} g) f .   
\end{equation}   
It is easy to show $gZ + Z^{*sT} g = 0$ so that $\delta (f^{\dag}~g~f) = 0$.   
If we  look at $U(2/1)$ parts   
\begin{equation}   
\delta F =   
 \left(   
\begin{array}{cc}   
i H  & 0 \\   
 0 &  i \omega   
\end{array}  \right)   
\end{equation}   
giving   
\begin{equation}   
\delta f = i H f, ~~~~~ \delta C = i \omega C   
\end{equation}   
and   
\begin{equation}   
\delta F^{\dag} = F^{\dag}   
 \left(   
\begin{array}{cc}   
- i H  & 0 \\   
 0 & - i \omega   
\end{array}  \right)   
\end{equation}   
giving   
\begin{equation}   
\delta f^{\dag} = - i f^{\dag} H, ~~~~~ \delta C^{*} = - i \omega C^{*} ,   
\end{equation}   
we obtain   
\begin{equation}   
\delta (f^{\dag} f) = 0 , ~~and~~~ \delta (C^{*} C) = 0.   
\end{equation}   
Similarly   
\begin{equation}   
\delta F =    
 \left(   
\begin{array}{cc}   
0  & \eta   \\   
 i \eta^{\dag} &  0   
\end{array}  \right)   
 F   
\end{equation}   
gives   
\begin{equation}   
\delta f = \eta C, ~~~~~\delta C = i \eta^{\dag} f,   
\end{equation}   
and using   
\begin{equation}   
\delta f^{\dag} = C^{*} \eta^{\dag}, ~~and~~   
\delta c^{*} = - i \eta^{T} f^{*} = i f^{\dag} \eta ,   
\end{equation}   
we arrive at    
\begin{equation}   
\delta (f^{\dag} f + i C^{*} C ) = 0   
\end{equation}   
or   
\begin{equation}   
\delta (i f^{\dag} f - C^{\dag} C )= 0   
\end{equation}   
Defining $\bar{f} = i f^{\dag}$, we have   
\begin{equation}   
\delta (\bar{f} f - C^{\dag} C) = 0.   
\end{equation}   
   
If we now define    
\begin{equation}   
h_{AB \dot{A} \dot{B}} = (\bar{X})_{AB} (x) (X)_{\dot{A} \dot{B}} (x)   
\end{equation}   
where $A,B = 0, 1, \ldots, 6$ as before,  then the subset $h_{0b0\dot{b}}$   
antisymmetric in the first and last pairs of   
indices would describe $q \bar{q}$ mesons, $h_{ab\dot{a} \dot{b}}$   
symmetric in first and last pairs of indices  would describe    
 $q^{2} \bar{q}^{2}$ exotic mesons,  $h_{0b\dot{a} \dot{b}}$ antisymmetric   
in the first and symmetric in the last pair of indices would describe   
$q^{2} q$ baryons, and $h_{ab 0 \dot{b}}$ symmetric in the first and   
antisymmetric in the last pair of indices would describe   
$\bar{q}^{2} \bar{q}$ antibaryons.   
   
Aside from $h_{AB \dot{A} \dot{B}}$ describing baryons, antibaryons,   
mesons and exotics, this algebra can be extended to include preons $F_{A}$,   
antipreons $F_{\dot{A}}$, $X_{AB}$ describing $(\bar{q}^{2} q)$, and   
$X_{\dot{A} \dot{B}}$ describing $(q^{2} \bar{q}$. Gauge bosons and    
gauginos can be in the adjoint representation $V_{A \dot{B}}$. We shall   
explore these aspects as well as building meson-baryon Lagrangians    
in another publication.   
   
We note that, since   
\begin{equation}   
(Z_{AB})^{*} = (Z^{*})_{\dot{A} \dot{B}}   
\end{equation}   
we have   
\begin{equation}   
\delta F_{A} = Z_{AB}~F_{B} ~~and~~~   
\delta F_{\dot{A}} = Z_{AB}^{*}~F_{\dot{B}}     
\end{equation}

\section{Particle multiplets including a giant supermultiplet}           
Multiplet $X$ that sits in the adjoint representation of $SU(6/21)$ given by  
\begin{equation}  
X = \left(  
\begin{array}{cc}  
M & B  \\  
\bar{B}  & N   
\end{array}   \right).  
\end{equation}  
Here $M$ and $N$ are mesons and exotics, and $B$ and $\bar{B}$ are fermions.  
The $M$ and $N$ are square matrices, and $B$ is a rectangular matrix.   
Specifically $M = 6 \times \bar{6}$, $B = 6 \times 21$,  
$\bar{B} = \bar{21} \times \bar{6}$, and $N = \bar{21} \times 21$. $M$ and  
$N$ are taken to be Hermitian. If we have three flavors the $SU(6)$ content   
of these matrices are $M = 1 + 35$ (negative parity), $N = 1 + 35 + 405$  
(positive parity), and $B = 56 + 70$ (positive parity).  The fundamental   
representation ${\bf F}$ is the color triplet  
\begin{equation}  
{\bf F} =   
\left(  
\begin{array}{c}  
{\bf Q}  \\  \bar{\bf D}  
\end{array}   \right)   =  
\left(  
\begin{array}{c}  
{\bf q}  \\  \bar{\bf q} \times \bar{\bf q}  
\end{array}   \right) ,  
\end{equation}  
with ${\bf q} = 6 \times 1$ and $\bar{\bf q} \times \bar{\bf q} = \bar{21}  
\times 1$. Now let $\Xi$ be the superalgebra element of $SU(6/21)$ ($SU(3)$   
singlet). $Xi$ is a color singlet given by  
\begin{equation}  
\Xi = \left(  
\begin{array}{cc}  
m & b  \\  
\bar{b}  & n   
\end{array}   \right)  
\end{equation}  
and the transformation law for the fundamental representation  
$[3, (6 + \bar{21})]$ is  
\begin{equation}  
\delta {\bf F} = \Xi {\bf F} =  
\left(  
\begin{array}{cc}  
m & b  \\  
\bar{b}  & n   
\end{array}   \right)  
\left(  
\begin{array}{c}  
{\bf Q}  \\  \bar{\bf D}  
\end{array}   \right)   =  
\left(  
\begin{array}{cc}  
m {\bf Q}  & b \bar{{\bf D}} \\  
\bar{b} {\bf Q} & n \bar{{\bf D}}   
\end{array}   \right) =  
\left(  
\begin{array}{c}  
\delta {\bf Q}  \\  \delta \bar{\bf D}  
\end{array}   \right),    
\end{equation}  
\begin{equation}  
\overline{\delta {\bf F}} = \overline{{\bf F}} \Xi  =  
(\bar{\bf Q} m + {\bf D} \bar{b} , \bar{{\bf Q}} b + {\bf D} n) =  
(\delta \bar{{\bf Q}}, \delta {\bf D})  
\end{equation}  
or in the index notation  
\begin{equation}  
\delta q_{\alpha}^{i} = m_{\alpha}^{\beta} q_{\beta}^{i} +  
b_{\alpha \beta \gamma} (\bar{D}^{i})^{\beta \gamma} ,    \label{eq:fr}  
\end{equation}  
\begin{equation}  
(\delta \bar{D}^{i})^{\beta \gamma} =   
\bar{b}^{\alpha \beta \gamma}q_{\alpha}^{i} +   
n_{\rho \sigma}^{\beta \gamma} (\bar{D}^{i})^{\rho \sigma} .  \label{eq:fs}  
\end{equation}  
On the other hand, the transformation law for the adjoint representation is  
\begin{equation}  
\delta \Xi = i [\Xi, X] ,  
\end{equation}  
where  
\begin{equation}  
\Xi X =  
\left(  
\begin{array}{cc}  
m & b  \\  
\bar{b}  & n   
\end{array}   \right)  
\left(  
\begin{array}{cc}  
M & B  \\  
\bar{B}  & N   
\end{array}   \right) =  
\left(  
\begin{array}{cc}  
m M + b \bar{B} & m B + b N \\  
\bar{b} M + n \bar{B}  & \bar{b} B + n N  
\end{array}   \right),  
\end{equation}  
\begin{equation}  
X \Xi =  
\left(  
\begin{array}{cc}  
M m + B \bar{b} & M b + B n \\  
\bar{B} m + N \bar{b}  & \bar{B} b + N  n  
\end{array}   \right) ,  
\end{equation}  
so that  
\begin{equation}  
\left(  
\begin{array}{cc}  
\delta M & \delta B  \\  
\delta \bar{B}  & \delta N   
\end{array}   \right) = i  
\left(  
\begin{array}{cc}  
[M, m] + b \bar{B} - \bar{B} b & M b - B n + b N - M b\\  
- \bar{B} m + n \bar{B} - N \bar{b} + \bar{b} M  &   
[n, N] + \bar{b} B - \bar{B} b   
\end{array}   \right) .  
\end{equation}  

Next we build a giant supermultiplet containing $M$, $N$, $L$, $B$,   
$\bar{B}$, $Q$, $\bar{Q}$, $D$, and $\bar{D}$. The fundamental representation  
of $U(6/21) \times [SU(3)^{c}]_{triplet}$ and the adjoint representation of  
$U(6/21) \times [SU(3)^{c}]_{singlet}$ fits in the adjoint representation of   
an octonionic version of $U(6/22)$ denoted by $Z$, given by   
\begin{eqnarray}  
Z  &  =  &  u_{0}   
\left(  
\begin{array}{ccc}  
M & B  &  0 \\  
B^{\dag}  &  N  & 0  \\  
0  &  0  &  0  
\end{array}   \right)   +  
u_{0}^{*}   
\left(  
\begin{array}{ccc}  
0 & 0  &  0 \\  
0 & 0  &  0  \\  
0  &  0  &  L      
\end{array}   \right)   +  
{\bf u} \cdot  
\left(  
\begin{array}{ccc}  
0 & 0  &  {\bf Q} \\  
0  &  0  &  {\bf D}^{*} \\  
0  &  0  &  0  
\end{array}   \right)   +   \nonumber  \\
   &      &
{\bf u}^{*}  \cdot  
\left(  
\begin{array}{ccc}  
0 & 0  &  0 \\  
0  &  0  & 0  \\  
\epsilon {\bf Q}^{\dag}  &  \epsilon {\bf D}^{T}  &  0  
\end{array}   \right)   \nonumber   \\  
   &      &  
 =  \left(  
\begin{array}{ccc}  
u_{0} M &  u_{0} B  &  {\bf u} \cdot {\bf Q} \\  
u_{0} B^{\dag}  &  u_{0} N  &  {\bf u} \cdot {\bf D}^{*}  \\  
\epsilon  {\bf u}^{*} \cdot {\bf Q}^{\dag}  &  \epsilon   
 {\bf u}^{*} \cdot {\bf D}^{T}  &   u_{0}^{*} L   
\end{array}   \right)     
\end{eqnarray}  
where mesons $M$ ($6 \times \bar{6}$) and exotics  $N$ ($\bar{21} \times 21$)  
are Hermitian; $B$ ($6 \times 21$), $\bar{B}$ ($\bar{21} \times \bar{6}$),  
$Q$ ($6 \times 1$), $\bar{Q}$ ($1 \times \bar{6}$),  
$D$ ($1 \times 21$), $\bar{D}$ ($\bar{21} \times 1$), and $L$ ($1 \times 1$);  
$\epsilon$ can be taken as $1$ if $u^{\dag} = \bar{u}^{*} = - u^{*}$, $-1$  
if $u^{\dag} =  u^{*}$, and zero. Closure properties of $Z$ matrices are  
such that  
\begin{equation}  
[Z, Z^{'}] = i Z^{''} ,  ~~~~~~ \{Z, Z^{'}\} =  Z^{'''}  
\end{equation}  
and in general they are nonassociative (Jacobian $J = f(Q,D) \neq 0$),   
except in the case when $\epsilon = 0$ we have  
\begin{equation}  
[[Z, Z^{'}] , Z^{''}] + [[Z^{'}, Z^{''}] , Z] + [[Z^{''}, Z] , Z^{'}] = 0 .  
\end{equation}  
Then we have a true superalgebra (non-semisimple) which is a contraction  
of a simple algebra that closes but does not satisfy the Jacobi identity. In   
both cases we get an extension of $U(6/21)$ considered by Miyazawa  
\cite{mi}.  
  
We now consider the element of the algebra   
\begin{equation}  
\Omega =   
\left(  
\begin{array}{ccc}  
u_{0} m &  u_{0} b  &  {\bf u} \cdot \mbox{\boldmath $\xi$} \\  
u_{0} b^{\dag}  &  u_{0} n  &  {\bf u} \cdot {\bf d}^{*}  \\  
\epsilon  {\bf u}^{*} \cdot \mbox{\boldmath $\xi$}^{\dag}  &  \epsilon   
 {\bf u}^{*} \cdot {\bf d}^{T}  &   u_{0}^{*} \ell   
\end{array}   \right)     
\end{equation}  
where  
\begin{equation}  
\left(  
\begin{array}{cc}  
u_{0} m &  u_{0} b   \\  
u_{0} b^{\dag}  &  u_{0} n  
\end{array}    \right)   
\end{equation}  
are the color singlet parameters [$U(6/21)$],  
\begin{equation}  
\left(  
\begin{array}{c}  
 {\bf u} \cdot \mbox{\boldmath $\xi$} \\  
{\bf u} \cdot {\bf d}^{*}   
\end{array}   \right) ~~~~~and~~~    
 ( \epsilon  {\bf u}^{*} \cdot \mbox{\boldmath $\xi$}^{\dag}~~~  
\epsilon  {\bf u}^{*} \cdot {\bf d}^{T} )  
\end{equation}  
are the colored parameters,    
\begin{equation}  
(u_{0} b ~~{\bf u} \cdot \xi)~~~ and~~~  
\left(  
\begin{array}{c}  
 u_{0} b {\dag}   \\  
\epsilon {\bf u}^{*} \cdot \mbox{\boldmath $\xi$}^{\dag}   
\end{array}   \right)  
\end{equation}  
are the fermionic parameters, and $m \in SU(6)$.  
  
The change in $Z$ is given by  
\begin{equation}  
\delta Z = [\Omega, Z]  
\end{equation}  
which leads to  
\begin{equation}  
\delta {\bf Q} =  m {\bf Q} - M \mbox{\boldmath $\xi$} + b {\bf D}^{*} - B 
{\bf d}^{*}  
+ { \bf \xi} L - {\bf Q} \ell ,  
\end{equation}  
and  
\begin{equation}  
\delta {\bf D}^{*} =  b^{\dag} {\bf Q} - B^{\dag} \mbox{\boldmath $\xi$} + 
n {\bf D}^{*} -   
N {\bf d}^{*} + { \bf d}^{*} L - {\bf D}^{*} \ell .  
\end{equation}  
  
The $U(6/21)$ subgroup is obtained by taking  
\begin{equation}  
\mbox{\boldmath $\xi$} = 0, ~~~~ {\bf d} = 0 ~~~~ \ell = 0  
\end{equation}  
so that  
\begin{equation}  
\delta {\bf Q} =  m {\bf Q} + b {\bf D}^{*}   
\end{equation}  
\begin{equation}  
\delta {\bf D}^{*} =  b^{\dag} {\bf Q} + n {\bf D}^{*}  
\end{equation}  
which, in index form, is equivalent to equations (\ref{eq:fr}) and  
(\ref{eq:fs}).   This subgroup is valid for a Hamiltonian describing   
$q(x_{1})$ and $\bar{q} (x_{2})$, $q(x_{1})$ and $D (x_{2}$),  
$\bar{D}(x_{1})$ and $q (x_{2)}$, $\bar{D}(x_{1})$ and $D (x_{2})$  
interacting through a scalar potential $V = b {\bf r}$ as we have seen 
earlier \cite{catg}.  
  
In general for $\frac{m}{2}$ flavors and $n = \frac{1}{2} m (m + 1)$,  
we have  
\begin{equation}  
  Z  =  \left(  
\begin{array}{ccc}  
u_{0} M &  u_{0} B  &  {\bf u} \cdot {\bf Q} \\  
u_{0} B^{\dag}  &  u_{0} N  &  {\bf u} \cdot {\bf D}^{*}  \\  
\epsilon  {\bf u}^{*} \cdot {\bf Q}^{\dag}  &  \epsilon   
 {\bf u}^{*} \cdot {\bf D}^{T}  &   u_{0}^{*} L   
\end{array}   \right)     
= \left(  
\begin{array}{ccc}  
m \times m &  m \times n  &  m \times 1  \\  
n \times m  &  n \times n  &  n \times 1  \\  
1 \times m  &  1 \times n  &  1 \times 1   
\end{array}   \right)   .  
\end{equation}  
  
As examples, for $2$ flavors, $M = 4 \times 4$, $N = 10 \times 10$; for  
$6$ flavors (icluding the top quark), $M= 12 \times 12$, $N = 78 \times  
78$.  
  
The automorhism group of this algebra includes  
$SU(m) \times SU(n) \times SU(3)^{c}$. If $m = 6$, it includes $ SU(6)  
\times SU(3)^{c}$. If ${\bf Q}$ is Majorana and ${\bf D}$ real, then the group  
becomes $Osp(n/m) \times SU(3)^{c}$, with subgroup $Sp(2n/R) \times  
O(m) \times SU(3)^{c}$. We shall explore quark models built by use of such  
groups as well as use of  
auxiliary octonions of quadratic norm $\frac{1}{2}$ which are related to  
the split octonion units we used in this paper in a subsequent publication  
\cite{scbb}.

\section{Relativistic formulation through the spin realization of the 
Wess-Zumino algebra}

It is possible to use a spin representation of the Wess-Zumino algebra 
to write first order relativistic equations for quarks and diquarks that are  
invariant under supersymmetry transformations. In this section we  
briefly deal 
with such Dirac-like supersymmetric equations. A discussion of  
experimental possibilities for the observation of the diquark structure 
and exotic $\bar{D}-D = (\bar{q}\bar{q})(qq)$ mesons
will be given elsewhere \cite{scpu} \cite{scbb}. For a very nice  
discussion of the experimental situation we refer the reader to a  
recent preprint by Anselmino, $et.al.$ \cite{anse}. Also, for a historical 
review of dynamical supersymmetries we refer the reader to a recent article by 
Iachell \cite{iacho}. 
 
There is a spin realization of the Wess-Zumino super-Poincar\'{e} algebra 
\begin{equation} 
[p_{\mu}, p_{\nu}] = 0 ,~~~~~[D_{\alpha}, p_{\mu}] = 0, 
\end{equation} 
\begin{equation} 
[\bar{D}_{\dot{\beta}}, p_{\mu}] = 0, ~~~~~ 
[D^{\alpha}, \bar{D}^{\dot{\beta}}] = \sigma_{\mu}^{\alpha \dot{\beta}} 
p^{\mu} 
\end{equation} 
with $p_{\mu}$ transforming like a $4$-vector and $D^{\alpha}$, 
$\bar{D}^{\dot{\beta}}$ like the left and right handed spinors under the  
Lorentz group with generators $J_{\mu \nu}$. 
 
We also note that 
\begin{equation} 
[J_{\mu \nu}, p_{\lambda}] = \delta_{\mu \nu} p_{\lambda} - 
\delta_{\nu \lambda} p_{\mu}, 
\end{equation} 
and 
\begin{equation} 
[J, J] = J . 
\end{equation}  
 
The finite non unitary spin realization is in terms of $4 \times 4$ 
matrices for $J_{\mu \nu}$ and $p_{\nu}$ 
\begin{equation} 
J_{\mu \nu} = \frac{1}{2} \sigma_{\mu \nu} = 
\frac{1}{4i} [\gamma_{\mu}, \gamma_{\nu}] , 
\end{equation} 
\begin{equation} 
J_{\mu \nu}^{L} = \frac{1-\gamma_{5}}{2} \frac{1}{2} \sigma_{\mu \nu} = 
\Sigma_{\mu \nu}^{L}   , 
\end{equation} 
\begin{equation} 
p_{\mu} = \Pi_{\mu}^{L} = \frac{1-\gamma_{5}}{2} \gamma_{\mu} .  \label{eq:va} 
\end{equation} 
 
Introducing two Grassmann numbers $\theta_{\alpha}$ ($\alpha = 1, 2$) 
that transforms 
like the components of a left handed spinor and commute with the Dirac 
matrices $\gamma_{\mu}$, we have the representation 
\begin{equation} 
D_{\alpha} = \Delta_{\alpha} = \frac{\partial}{\partial \theta_{\alpha}}, 
\end{equation} 
\begin{equation} 
\bar{D}^{\dot{\beta}} = \bar{\Delta}^{\dot{\beta}} = \theta_{\alpha} 
\sigma_{\mu}^{\alpha \dot{\beta}} \Pi_{\mu}^{L} . 
\end{equation} 
 
Such a representation of the super-Poincar\'{e} algebra acts on a Majorana  
chiral superfield 
\begin{equation} 
S(x, \theta) = \psi (x) + \theta_{\alpha} B^{\alpha} (x) + \frac{1}{2} 
\theta_{\alpha} \theta^{\alpha} \chi (x) . 
\end{equation} 
 
Here $\psi$ and $\chi$ are Majorana superfields associated with fermions and 
$B^{\alpha}$ has an unwritten Majorana index and a chiral spinor index 
$\alpha$, so that it represents a boson. 
 
Note that the sum of the two representations we wrote down is also a realization 
of the Wess-Zumino algebra. 
 
On the other hand we have the realization of $p_{\mu}$ in terms of the  
differential operator $-i \partial_{\mu} = -i \frac{\partial}{\partial  
x^{\mu}}$. In the Majorana representation, the operator $\gamma_{\mu} 
\partial_{\mu} = i \gamma_{\mu} p_{\mu}$ is real, and $\psi  = \psi ^{c} = 
\psi ^{*}$.Let us now define $\psi_{L}$ and $\psi_{R}$ by 
\begin{equation} 
\psi_{L} = \frac{1}{2} (1 + \gamma_{5}) \psi,        
\end{equation} 
and 
\begin{equation} 
\psi_{R} = \frac{1}{2} (1 - \gamma_{5}) \psi = \psi_{L}^{*}  . 
\end{equation} 
The free particle Dirac equation can now be written as 
\begin{equation} 
\Pi_{\mu}^{L} \partial_{\mu} \psi_{L} = m \psi_{L}^{*} ,   \label{eq:pi} 
\end{equation} 
or 
\begin{equation} 
\Pi_{\mu} p^{\mu} \psi_{L} = - i m \psi_{L}^{*}. 
\end{equation} 
We can introduce 
\begin{equation} 
S_{L} = \frac{1}{2} (1 + \gamma_{5}) S , 
\end{equation} 
\begin{equation} 
S_{R} = \frac{1}{2} (1 - \gamma_{5}) S = S_{L}^{*}. 
\end{equation} 
Then equation (\ref{eq:pi}) generalizes to the superfield equation 
\begin{equation} 
\Pi_{\mu}^{L} \partial_{\mu} S_{L} = m S_{L}^{*} .   \label{eq:pia} 
\end{equation} 
or 
\begin{equation} 
\Pi_{\mu} p^{\mu} S_{L} = - i m S_{L}^{*}. 
\end{equation} 
Now consider the supersymmetry transformation 
\begin{equation} 
\delta S_{L} = (\xi^{\alpha} \Delta_{\alpha} + \bar{\xi}_{\dot{\beta}} 
\bar{\Delta}^{\dot{\beta}}) S_{L} = \Xi S_{L} .   \label{eq:pib} 
\end{equation} 
This transformation commutes with the operator $\Pi_{\mu}^{L} \partial_{\mu}$ 
so that 
\begin{equation} 
\Pi_{\mu}^{L} \partial_{\mu} (S_{L} + \delta S_{L}) = m 
(S_{L} + \delta S_{L})^{*} . 
\end{equation} 
 
If $\psi_{L}$ is a left handed quark and $B^{\alpha}(x)$ an antidiquark 
with the same mass as the quark, equation (\ref{eq:pib}) provides a  
relativistic form of the quark antidiquark symmetry which is in fact broken 
by the quark-diquark mass difference. The scalar supersymmetric potential  
is introduced through $m \longrightarrow m + V_{s}$ as before and the equation  
(\ref{eq:pia}) remains supersymmetric. By means of this formalism, it is 
possible to reformulate the treatments given in the earlier sections in  
first order relativistic form. 
 
To write equations in the first order form, we consider $V$ and $\Phi$ 
given in terms of the boson fields by 
\begin{equation} 
V = i \gamma_{\mu} V_{\mu} -\frac{1}{2}   \sigma_{\mu \nu} 
\end{equation} 
and 
\begin{equation} 
\Phi = i \gamma_{5} \phi + i \gamma_{5} \gamma_{\mu} \phi_{\mu} . 
\end{equation} 
In the Majorana representation  
\begin{equation} 
V^{*} = - V , ~~~~~and ~~~~~~ \Phi = \Phi^{*}. 
\end{equation} 
 
We now define the left and right handed component fields by 
\begin{equation} 
V_{L} = \frac{1- \gamma_{5}}{2} V, ~~~and~~~~V_{R}= \frac{1+\gamma_{5}}{2} V 
so that\end{equation} 
\begin{equation} 
V_{R} = - V_{L}^{*} 
\end{equation} 
with 
\begin{equation} 
V_{L} = i \frac{1- \gamma_{5}}{2} \gamma_{\mu} V_{\mu} - 
 \frac{1- \gamma_{5}}{2} \frac{1}{2} \sigma_{\mu \nu} V_{\mu \nu} 
\end{equation} 
and using equation (\ref{eq:va}) we have 
\begin{equation} 
V_{L} = i \Pi_{\mu}^{L} V_{\mu} - 
 \frac{1- \gamma_{5}}{2} \frac{1}{2} \sigma_{\mu \nu} V_{\mu \nu}. 
\label{eq:vi} 
\end{equation} 
 
Noting that for any $a_{\mu}$ and $b_{\nu}$ 
\begin{equation} 
\Pi_{\mu} \Pi_{\nu}^{*} a_{\mu} b_{\nu} =  
\frac{1- \gamma_{5}}{2} \gamma_{\mu} \frac{1+ \gamma_{5}}{2} \gamma_{\nu} 
a_{\mu} b_{\nu} = \frac{1- \gamma_{5}}{2} \gamma_{\mu} \gamma_{\nu} 
a_{\mu} b_{\nu} 
\end{equation} 
so that using the definition of $\sigma_{\mu \nu}$ we have 
\begin{equation} 
\frac{1}{2} (\Pi_{\mu} \Pi_{\nu}^{*} - \Pi_{\nu} \Pi_{\mu}^{*}) = 
\frac{1- \gamma_{5}}{2} i \sigma_{\mu \nu}. 
\end{equation} 
Incorporating this in equation (\ref{eq:vi}) leads to 
\begin{equation} 
V_{L} = i \Pi_{\mu}^{L} V_{\mu} + \frac{i}{2}  
(\Pi_{\mu} \Pi_{\nu}^{*} - \Pi_{\nu} \Pi_{\mu}^{*}) V_{\mu \nu} . 
\label{eq:vl} 
\end{equation} 
Letting $\Sigma_{\mu \nu}^{L} = \Pi_{\mu} \Pi_{\nu}^{*} -  
\Pi_{\nu} \Pi_{\mu}^{*}$, equation (\ref{eq:vl}) becomes 
\begin{equation} 
V_{L} = i \Pi_{\mu}^{L} V_{\mu} + \frac{i}{2} \Sigma_{\mu \nu}^{L} V_{\mu \nu} . 
\end{equation} 
So, we now have a first order equation 
\begin{equation} 
\Pi_{\mu}^{L} \partial_{\mu} V_{R} = m V_{L}.   \label{eq:vc} 
\end{equation} 
Also, 
\begin{equation} 
\Pi_{\mu}^{R} \partial_{\mu} V_{L} = m V_{R}. 
\end{equation} 
Therefore 
\begin{equation} 
\Pi_{\mu}^{L} \Pi_{\mu}^{R} \partial_{\mu} \partial_{\nu} V_{L} =  
 m \Pi_{\mu}^{L} \partial_{\mu} V_{R} = m^{2} V_{L} 
\end{equation} 
which after substitution of equation (\ref{eq:vc}) gives 
\begin{equation} 
\Box{V_{L}} = m^{2} V_{L}. 
\end{equation} 
Similarly, for the $\Phi$ part we can write 
\begin{equation} 
\Pi_{\mu}^{L} \partial_{\mu} \Phi_{R} = m \Phi_{L} 
\end{equation} 
again, as before 
\begin{equation} 
\Phi_{L} = \frac{1- \gamma_{5}}{2} \Phi, ~~~~~~and~~~~~ 
\Phi_{R} = \frac{1+ \gamma_{5}}{2} \Phi 
\end{equation} 
with  
\begin{equation} 
\Phi_{R} = \Phi_{L}^{*}. 
\end{equation} 
We can then repeat the above procedure for the $\Phi$ fields.

\section{Acknowledgments}
Stimulating discussions with I. Giannakis, F. Iachello, R. Khuri, P. Orland
and H.C. Tze are gratefully acknowledged.  
This work was supported in part by DOE contracts No. 
DE-AC-0276 ER 03074 and 03075; NSF Grant No. DMS-8917754; and PSC-CUNY 
Research Awards.

\end{document}